\documentclass[noshowpacs,amsmath,twocolumn,
aps,prb]
{revtex4-1}

\usepackage{setspace}
\usepackage{amsmath}
\usepackage{graphicx}
\usepackage[nearskip,margin = 0pt]{subfig}
\usepackage{subfig}

\usepackage{verbatim}
\usepackage{amsfonts}
\usepackage{amssymb}
\usepackage{epstopdf} 
\usepackage{xcolor}
\usepackage{color}
\DeclareGraphicsExtensions{.pdf,.eps,.png,.jpg,.mps} 
\usepackage{ragged2e}
\usepackage{hyperref}
\hypersetup{
    colorlinks=true,
    linkcolor=blue,
    citecolor=blue,
    filecolor=blue, 
    urlcolor=blue,
}
\usepackage{mathrsfs}

\begin{document}

\title{Observation of spatiotemporal stabilizer in a multi-mode fibre laser}

\author{Chenxin Gao,$^{1}$  Chengjiu Wang,$^{1}$ Zhenghao Jiao,$^{1}$ Bo Cao,$^{1}$ Xiaosheng Xiao,$^{2}$ Changxi Yang,$^{1}$ and Chengying Bao,$^{1}$$^\dag$ \\
$^1$State Key Laboratory of Precision Measurement Technology and Instruments, Department of Precision Instruments, Tsinghua University, Beijing 100084, China.\\
$^2$State Key Laboratory of Information Photonics and Optical Communications, School of Electronic Engineering, Beijing University of Posts and Telecommunications,  Beijing 100876, China.\\
Corresponding authors: $^\dag$cbao@tsinghua.edu.cn
}

\maketitle
\newcommand{\ts}{\textsuperscript}

\newcommand{\tsb}{\textsubscript}

%%%%%%%%%%%%%%%%%%%%%%%% Main Text %%%%%%%%%%%%%%%%%%%%%%%

{\bf Spatiotemporal mode-locking (STML) has become an emerging approach to realize organized wavepackets in high-dimensional nonlinear photonic systems. Mode-locking in one dimensional systems employs a saturable absorber to resist fluctuations in the temporal domain. Analogous suppression of fluctuations in the space-time domains to retain a consistent output should also exist for STML. However, experimental evidence of such a resistance remains elusive, to our knowledge. Here, we report experimental observation of such a spatiotemporal stabilizer in STML, by embedding a spatial light modulator (SLM) into a multi-mode fibre (MMF) laser. Mode decomposition reveals the mode content remains steady for an STML state when applying phase perturbations on the SLM. Conversely, the mode content changes significantly for a non-STML lasing state. Numerical simulations confirm our observation and show that spatial filtering and saturable absorber mainly contribute to the observed stability. The capability to resist the spatial phase fluctuations is observed to depend on the intracavity pulse energy as well as the modal pulse energy condensed in the low-order modes. %by varying the offset splicing between the MMFs. 
Our work constitutes another building block for the concept of STML in multi-mode photonic systems.}

\begin{figure*}[th!]
\includegraphics[width=0.95\linewidth]{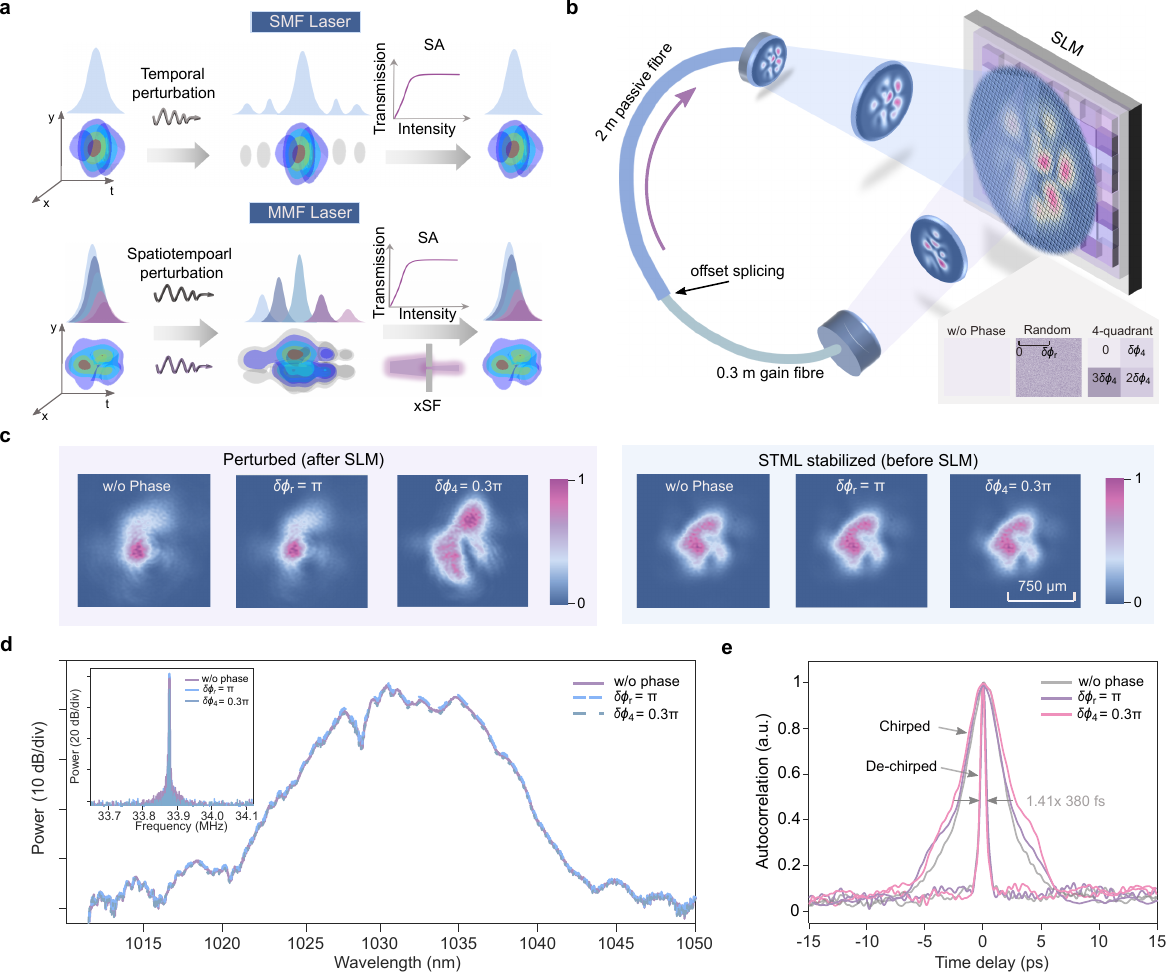}
\centering
\captionsetup{singlelinecheck=off, justification = RaggedRight}
\caption{{\bf Spatiotemporal mode-locking (STML) in a multi-mode fibre (MMF) laser with spatial perturbation.} {\bf a} Temporal mode-locking in single mode lasers resists fluctuations in the background by a saturable absorber (SA). For STML, spatial filtering (xSF) and SA act as a spatiotemporal stabilizer against perturbations to retain a steady state. {\bf b} Layout  of the MMF laser with a spatial light modulator (SLM) included in the cavity. We used two types of spatial phase, two-dimension random phase and 4-quadrant phase, to perturb the STML laser. {\bf c} Measured beam profiles right after the SLM and after circulating the cavity (before SLM). The laser beam changes considerably after loading the spatial phase perturbation, but the beam profiles stay nearly the same after propagation in the cavity. {\bf d} Optical spectra with different spatial phase written on the SLM. The inset shows the measured repetition rate signals.
e Autocorrelation traces of the output pulses before and after compression by a grating pair with/without spatial perturbation.
}
\label{fig1}
\end{figure*}

\noindent{\bf Introduction.} Light propagation in multi-mode photonic systems is receiving increasing attention recently \cite{cao2019complex,Wright_NP2022physics}. For example, the capability to generate highly distorted speckles in multi-mode fibres (MMFs) opens the door to various applications such as compact imagers and spectrometers \cite{cao2023controlling}. 
Conversely, the formation of organized nonlinear wavepackets in MMFs contributes to advancing our comprehension of light thermodynamics \cite{Christodoulides_Science2023observation,Wise_NP2022direct}, multi-mode soliton dynamics \cite{Wise_NC2013,Wise_NP2015controllable}, nonlinear beam self-cleaning \cite{Krupa_NP2017spatial,Wise_OL2016kerr}, and facilitates the generation of spatiotemporal mode-locking (STML) pulses \cite{Wright_NP2022physics,wright2017spatiotemporal,Bao_LSA2023spatiotemporal,Yang_OL2018observation,teugin2019spatiotemporal,2021PRL_ding,guo2021real,Huang_NC2022synthesized,gao2023all}. STML enables optical phase synchronization in both temporal and spatial domains, so that the large mode contents of MMFs can be unlocked for high energy pulse generation \cite{wright2017spatiotemporal,Bao_LSA2023spatiotemporal,Wise_OL2022multimode,Bao_OL2022self,teugin2019spatiotemporal,2021PRL_ding}. These STML MMF lasers are also refreshing our knowledge of fundamental laser physics \cite{wright2020mechanisms,haken1986laser}. As a requirement for effective mode-locking in single-mode systems, fluctuations in the temporal background should be suppressed by a saturable absorber (SA, see Fig. \ref{fig1}a), to prevent noise from devastating the mode-locking \cite{Haus_APB1997ultrashort,Haus_STJQE2000mode}. It is straightforward to ask whether a similar spatiotemporal stabilizer against spatial fluctuations exists for STML (Fig. \ref{fig1}a). Although the ability to retain STML against perturbations has been demonstrated since the birth of STML \cite{wright2017spatiotemporal}, experimental evidence of spatiotemporal stabilization into a same state against peturbations remains elusive, to our knowledge. Moreover, the spatial complexity in multi-mode systems endows strong spatiotemporal interactions including spatiotemporal saturable absorption, modal gain competition and mode-dependent dissipation (including spatial filtering, abbreviated as xSF to distinguish from spectral filtering, $\omega$SF) \cite{2021PRL_ding,wright2020mechanisms,wright2017multimode,Bao_LSA2023spatiotemporal}, which have no counterparts in single-mode systems. Therefore, it is critical to understand the roles of different effects in underlying the potential spatiotemporal stabilizer.

In this work, we report the experimental and numerical observation of such a spatiotemporal stabilizer in an STML MMF laser (Fig. \ref{fig1}a). We embedded a spatial light modulator (SLM) into an MMF laser for our experiment (Fig. \ref{fig1}b). SLM in such a laser configuration has been proposed for spatial mode control in STML lasers \cite{fu2018several} and has been experimentally used for initializing STML in an MMF laser \cite{wei2020harnessing}. Here, we used it to write different spatial phase on the laser beam so as to control the spatial fluctuations for STML pulses quantitatively. With spatial phase perturbations loaded, the laser beam profile right after the SLM changes substantially, but the change is rectified after propagating in the MMF laser cavity. Mode decomposition based on a convolution neural network (CNN) %\textcolor{red}{together with the stochastic parallel gradient descent (SPGD) algorithm} 
method  \cite{an2020deep,bruning2013comparative,an2019learning,an2020deep} shows the mode content stays steady with varying random phase on the SLM, while the content changes randomly for a non-STML lasing state. Numerical simulations based on the generalized multi-mode nonlinear Schr$\rm\ddot{o}$ndinger equations (GMNLSEs) \cite{wright2017multimode,zhang2022all,zhang2023investigation,Bao_LSA2023spatiotemporal} and the attraction dissection theory \cite{wright2020mechanisms} both confirm the existence of such a spatiotemporal stabilizer and reveal that xSF and SA are critical to this stabilization. By changing the splicing offset between the MMFs, we observe the capability of the spatiotemporal stabilizer to resist the random phase perturbation increases with pulse energy, as well as the modal energy condensed in the low-order modes. Our work firmly strengthens the concept of STML and quantify the resistance to spatial fluctuations in STML. 
%, which contributing to a better understanding and implementation of stable STML fibre lasers.

\begin{figure*}[th!]
\includegraphics[width=\linewidth]{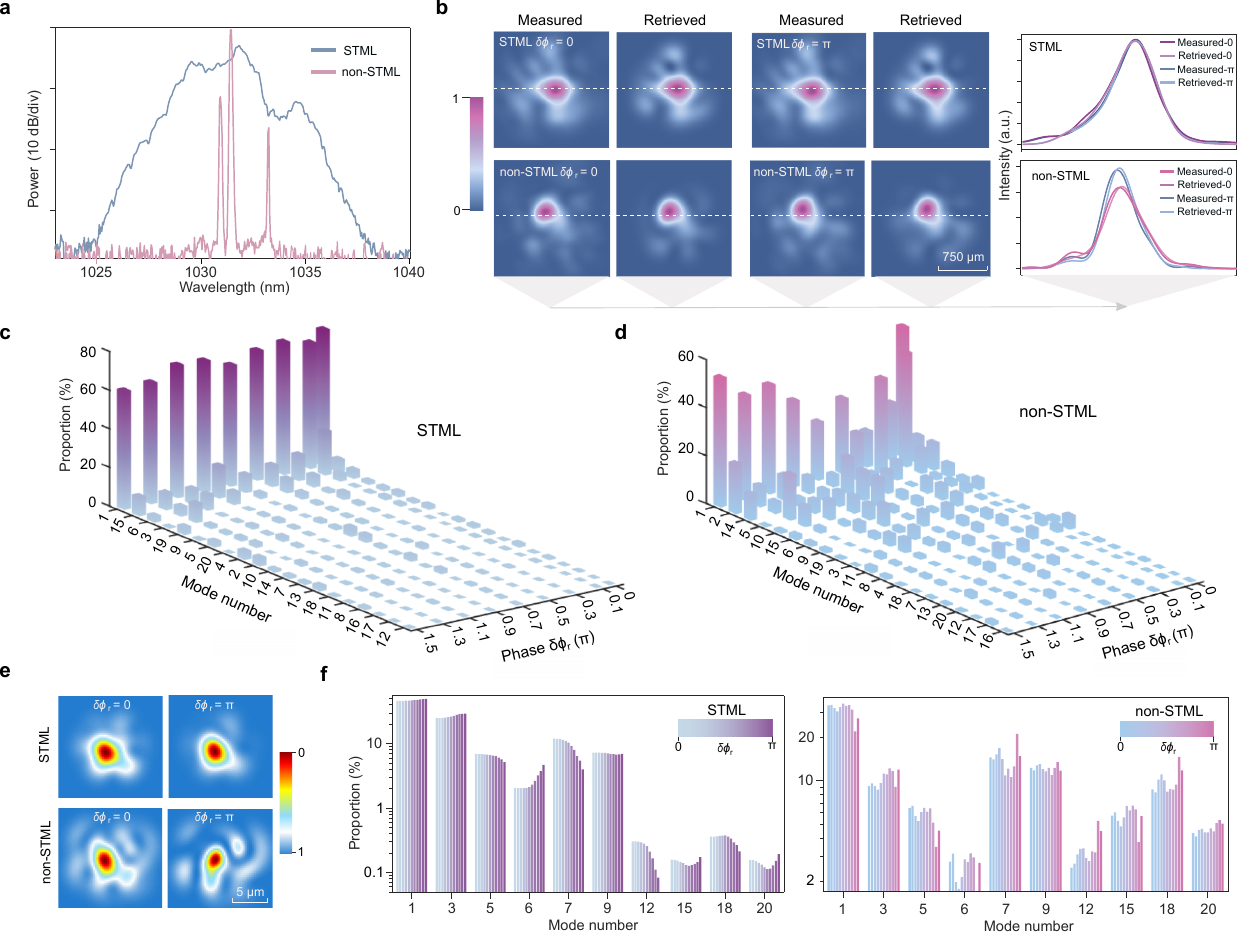}
\centering
\captionsetup{singlelinecheck=off, justification = RaggedRight}
\caption{ {\bf Spatial mode stability in spatiotemporal mode-locking (STML).} 
{\bf a} Optical spectra for  an STML state and a non-STML state. {\bf b} Measured and retrieved beam profiles with $\delta \phi_r$=0, $\pi$ in the STML state and the non-STML states. The right panel shows the beam profiles along the dashed slices. {\bf c} Retrieved mode proportion under different $\delta \phi_r$ is steady for the STML state. {\bf d} Mode proportion under different $\delta \phi_r$ changes randomly for the non-STML state. 
{\bf e} Simulated beam profiles for an STML and a non-STML state with $\delta \phi_r$=0, $\pi$. 
{\bf f} Mode proportion with different random phase $\delta \phi_r$ for an STML state and a non-STML state in the simulation. }
\label{fig2}
\end{figure*}

\begin{figure*}[th!]
\includegraphics[width=\linewidth]{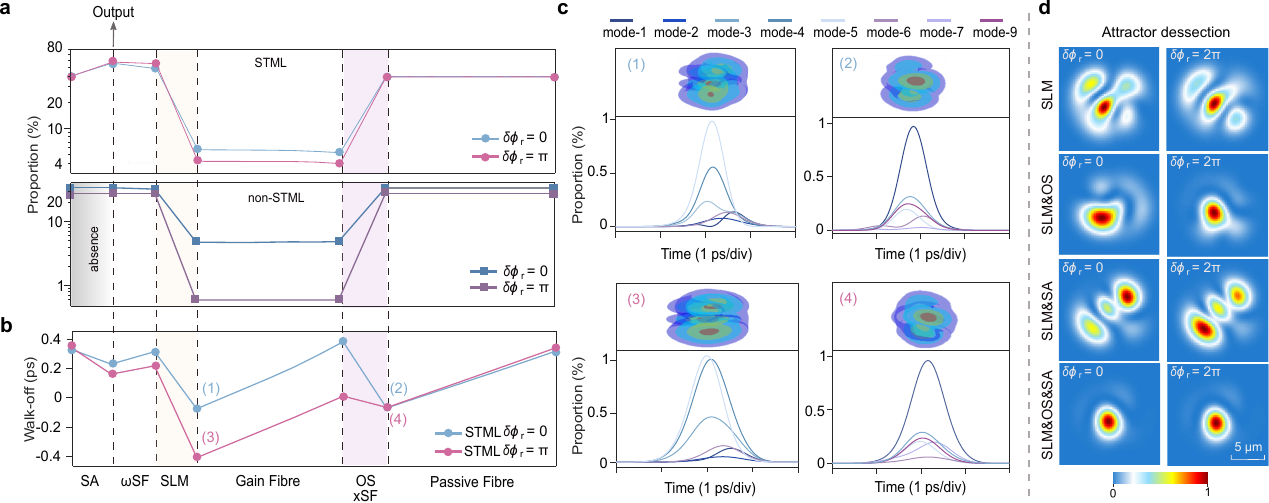}
\centering
\captionsetup{singlelinecheck=off, justification = RaggedRight}
\caption{ {\bf Intracavity spatiotemporal stabilizer propagation dynamics in simulations.} 
{\bf a} Intracavity change of power proportion for mode-1 with $\delta \phi_r$=0 and $\pi$ in an STML and a non-STML state. %The insets are the output beam profiles. 
The saturable absorber (SA) was not included in the simulation for the non-STML state. OS, offset splicing. {\bf b} Intracavity change in inter-mode walk-off between mode-1 and mode-3 with $\delta \phi_r$=0 and $\pi$ in an STML state.  
{\bf c} Modal pulses at representative positions (1-4) indicated in panel b.  
{\bf d} Output beam profiles simulated by the attractor dessection theory \cite{wright2020mechanisms}. From top to bottom, the images show the beam profiles if the laser dominated by the SLM only, SLM \& OS, SLM \& SA, and SLM \& OS \& SA operators with $\delta \phi_r$=0 or 2$\pi$. Beam stability was simulated with both OS and SA included.}
\label{fig3}
\end{figure*}

\noindent{\bf Observation of spatial stability in STML.} We built an MMF laser as illustrated in Fig. \ref{fig1}b. The cavity has a free-spectral-range of 33.9 MHz; the gain fibre is a 0.3 m long step-index (STIN) MMF Yb-doped fibre, while the passive fibre is a 2 m graded-index (GRIN) MMF supporting about 65 LP-modes at 1030 nm. These two MMFs were spliced with an offset of 8 $\mu$m. The offset splicing (OS) induces mode-dependent dissipation and contributes to xSF \cite{wright2017spatiotemporal,Bao_LSA2023spatiotemporal}. In addition, coupling from the free space into the STIN gain fibre also contributes to xSF, which only supports less than 6 modes. More details of the laser can be found in Supplementary Fig. S1. 

STML in the laser was first initiated by nonlinear polarization rotation (NPR) \cite{Haus_OL199377,Haus_APB1997ultrashort}, with the SLM loaded with a uniform phase. The output pulse has an energy of 7.1 nJ under a pump power of 4.9 W. The relatively low lasing efficiency is because we deliberately increased the cavity loss to avoid damaging the SLM. The output beam profile and optical spectrum are plotted in Fig. \ref{fig1}c and Fig. \ref{fig1}d, respectively. The corresponding 33.9 MHz repetition rate signal has a signal-to-noise ratio of 80 dB with a resolution bandwidth of 5 Hz (inset of Fig. \ref{fig1}d). We further measured the STML pulse width by an intensity autocorrelator. The direct output has a chirped pulse width of 2.9 ps, and can be compressed  to about 380 fs by a grating pair with a group delay dispersion of about $-$0.6 ps$^2$ (gray curves in Fig. \ref{fig1}e). %\textcolor{magenta}{The group delay dispersion of the grating pair is about -30.968 ps$^2$}. 

Then we loaded phase perturbations on the SLM. Two types of phase profiles were used; one is a two-dimension random phase with a uniform distribution between 0 and $\delta \phi_r$ and the other is a 4-quadrant phase varying from 0 to 3$\delta \phi_4$ (Fig. \ref{fig1}b). With the SLM phase perturbation loaded, the laser beam profile right after the SLM changes substantially (see Fig. \ref{fig1}c with $\delta \phi_r$=$\pi$ and $\delta \phi_4$=0.3$\pi$). Interestingly, after propagation in the cavity, the beam profiles before the SLM were observed to be nearly the same with the one without SLM phase perturbation (Fig. \ref{fig1}c). In other words, the STML laser has an ability to retain spatial stability against spatial perturbation. 

In the frequency domain, the perturbed optical spectra as well as the repetition frequency were also observed to be the same with those in the absence of perturbation (Fig. \ref{fig1}d). In the time domain, the perturbed STML pulse can still be compressed to about 380 fs, without adjusting the grating pair (Fig. \ref{fig1}e). The pulse compression measurement also suggests the STML pulse retains strong phase coherence in the presence of the spatial phase perturbation. As an aside, spatial sampling and spectral filtering measurements \cite{wright2017spatiotemporal} also show the laser outputs remain steady with the SLM phase perturbation loaded (Supplementary Sec. 2). Similar STML stability was also observed in an all STIN MMF laser \cite{gao2023all} with a much larger intermode dispersion (Supplementary Sec. 3).

\noindent \textbf{Mode decomposition analysis.} To quantify the STML stability, we employed a CNN approach augmented with the stochastic parallel gradient descent (SPGD) algorithm  \cite{an2019learning,an2020deep,bruning2013comparative} to investigate the mode content of the STML output (Methods). This method retrieves the mode content using the beam profile only and is relatively simple, compared to other mode decomposition methods for STML lasers \cite{wright2020mechanisms,Bao_LSA2023spatiotemporal}. As a caveat, this method should be used with care for STML pulses with broad spectra, see Supplementary Sec. 4. Here, we used the output after the 3 nm intracavity filter with a relatively narrow spectrum for mode analysis. %(see Supplementary Sec. 4 for details). 

The STML optical spectra for this measurement is shown in Fig. \ref{fig2}a (measured before the intracavity filter). The measured and retrieved beam profiles with $\delta \phi_r$=0 and $\pi$ in the STML state are plotted in Fig. \ref{fig2}b. Good agreement between the measured and retrieved beam profiles is observed. The right panel of Fig. \ref{fig2}b shows a horizontal slice of the beam along the dashed line. This agreement not only validates our mode decomposition method, but also confirms the stability of the output beam.

In contrast, when adjusting the waveplates to have a non-STML state (see Fig. \ref{fig2}a for the optical spectrum), considerable change of the output beams was observed (Fig. \ref{fig2}b). The intensity profiles along the dashed slice clearly show such variations. Note that the curves were normalized by the global beam peak intensity. Therefore, both the lineshape and the intensity differences show the change in the beam profile.

The retrieved mode content under different $\delta \phi_r$ for the STML and non-STML are plotted in Figs. \ref{fig2}c, d. For the STML state, the output beam is primarily dominated by the fundamental mode, since the splicing offset between the gain and GRIN MMFs is relatively small (8 $\mu$m). With an increasing $\delta\phi_r$, the fundamental mode proportion stays nearly the same (changes within [62\%, 70\%]). High-order modes (e.g., mode-15) may change stronger, but their changes do not impact the STML pulse significantly, as the absolute proportion is low. Similar mode content stability was observed for the 4-quadrant spatial phase perturbation, see Supplementary Sec. 5. Conversely, the retrieved mode content changes randomly for the non-STML state. For example, the fundamental mode proportion changes randomly within [15\%, 51\%] when increasing $\delta\phi_r$. This strong variation also confirms the random phase on the SLM can induce mode-dependent dissipation for the MMF laser, and rules out the possibility that the observed spatial mode stability in STML is caused by plain mode-independent-loss. %(\textcolor{magenta}{see also Supplementary Sec. 9}).

We further validated our observation by numerical simulations. The simulation model resembles those used in previous STML reports \cite{wright2017multimode,wright2017spatiotemporal,2021PRL_ding,gao2023all,Bao_LSA2023spatiotemporal}, but was augmented by an SLM (see Supplementary Sec. 6 for details). To reduce the computation burden, we selected 10 modes out of the 20 low-order modes for simulation (similar mode number reduction was also used in refs.\cite{2021PRL_ding,Wright_NP2022physics,liu2021buildup}). Figure \ref{fig2}e shows the simulated beam profile remains steady for an STML state when a 2-dimension random phase was added in the simulation. Nevertheless, the beam profile changes strongly for a non-STML state, when increasing $\delta \phi_r$ from 0 to $\pi$. The simulated mode content for STML and non-STML states with varying $\delta \phi_r$ is summarized in Fig. \ref{fig2}f. Consistent with the experiments, the simulated content remains steady for the STML state with a varying $\delta\phi_r$, but changes randomly for the non-STML state. 

\begin{figure*}[th!]
\centering\includegraphics[width=\linewidth]{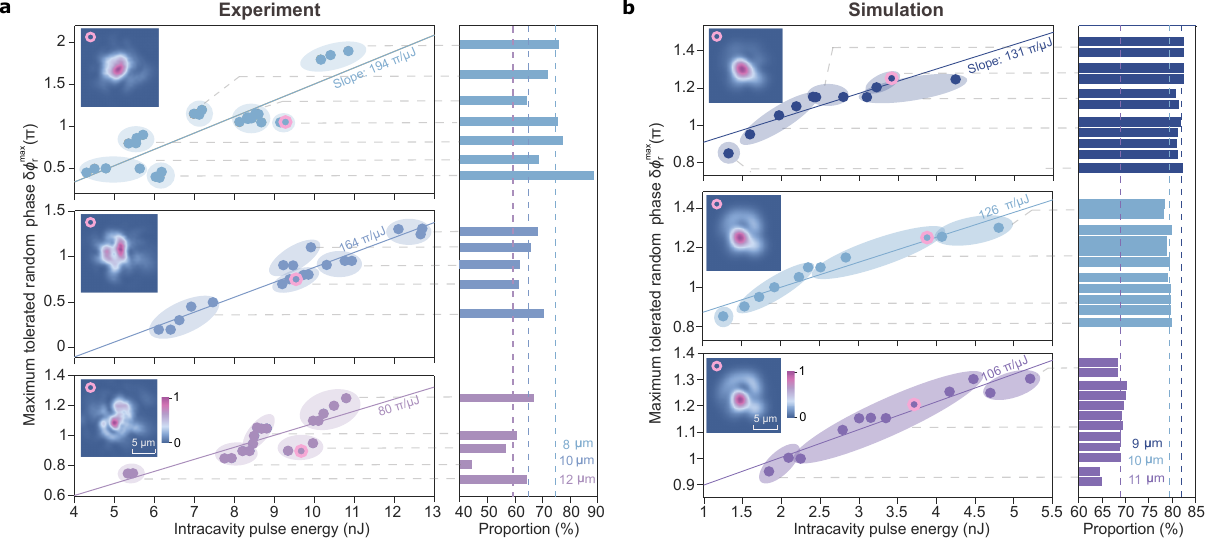}
\captionsetup{singlelinecheck=off, justification = RaggedRight}
\caption{{\bf Capability of the spatiotemporal stabilizer.} {\bf a,} The maximum random phase $\delta \phi_r^{\rm max}$ that can be tolerated by the STML laser under different lasing conditions. $\delta \phi_r^{\rm max}$ increases with increasing intracavity pulse energy. When changing the splicing offset from 8 $\mu$m to 10 $\mu$m and 12 $\mu$m, the slope of $\delta \phi_r^{\rm max}$ versus pulse energy decreases. The right panels show the low-order (mode-1 to mode-5) mode proportion for the STML states. {\bf b,} Simulated $\delta \phi_r^{\rm max}$ under different intracavity energy and splicing offset, showing similar trends to the experiment. The insets show the measured and simulated output beams that correspond to the dots with pink edges. The dashed lines in the right panels are the average proportion for a given splicing offset. %The output beams become more structured with increasing offset. %Each circle in the figure represents an STML state obtained by adjusting the polarization or coupling condition of the laser. 
}
\label{fig4}
\end{figure*}

\noindent{\bf Roles in the spatiotemporal stabilizer.} Simulations further enable us to analyze the roles of different laser components in underpinning the spatiotemporal stabilizer. We plot the intracavity mode proportion change for mode-1 under $\delta\phi_r$=0 and $\delta\phi_r$=$\pi$ in STML and non-STML states in Fig. \ref{fig3}a. Two major differences can be observed. One is that nonzero $\delta\phi_r$ increases the loss for mode-1 when coupling into the gain fibre from the SLM, and this loss increase is more significant for the non-STML state. The other is that the OS between the gain fibre and the passive fibre guides power back into mode-1, but only the STML state returns to the unperturbed level with $\delta \phi_r$=0.

%As shown in Fig. \ref{fig3}b, When there is a nonzero $\delta\phi_r$ at STML state, the proportion for mode-1 changes slightly at the output port. The S   LM induces larger loss than without random phase loaded. This extra loss receive some compensation at the offset splicing region when coupling from gain fibre to the passive MMF; more energy is guided back to mode-1. In addition, the SA induces smaller loss than without random phase loaded to compensation the change from phase disturbance.   

As a characteristics of STML, spatial perturbation from the SLM also penetrates into temporal perturbation, resulting in modal pulse walk-off (Fig. \ref{fig3}b). A walk-off change up to 0.6 ps is caused by the SLM with $\delta \phi_r$=$\pi$. It is also the OS between the gain fibre and the passive fibre that counteracts the large walk-off change and bind the modal pulses back to the unperturbed state. Similar temporal correction was simulated for other spatial modes, see the 3D pulses and the modal pulses in Fig. \ref{fig3}c, which correspond to positions (1-4) labelled in Fig. \ref{fig3}b. The SLM induces considerable changes for the modal pulses in both intensity and pulse timing (see positions (1, 3)). These changes also suggest the SLM induces mode-dependent-loss by simulations. Since the OS counteracts the walk-off, modal pulses at positions (2, 4) become similar again. The role of xSF in compensating systematic intermode walk-off has been well-recognized  \cite{wright2017spatiotemporal,gao2023all}. Here, we show that xSF also resists the spatial phase perturbation induced random temporal fluctuations. 

The existence of the spatiotemporal stabilizer can also be understood from the lowest loss principle for lasers \cite{wright2020mechanisms,haken1986laser}. The unperturbed STML pulses (e.g., pulse in Fig. \ref{fig3}c(2)) enjoy the lowest loss in the MMF laser system. The corresponding dissipative soliton can be understood as a nonlinear attractor in the system \cite{wright2020mechanisms,grelu2012dissipative,Wise_PRA2008dissipative}. Thus, the perturbed STML pulses tend to adjust themselves to approach the unperturbed one (e.g., pulse in Fig. \ref{fig3}c(4)), as long as the perturbation is relatively weak. Here, we elucidate how the STML state is spatiotemporally stabilized into the attractor in the presence of spatial perturbation.

To verify the above picture, we followed the attractor dissection theory \cite{wright2020mechanisms} to analyze the roles of different cavity components (see Supplementary Sec. 7 for details). Each laser cavity component is simplified as an operator in this model. Although STML dynamics is influenced by many operators, some operators dominate. Then, we can simplify the pulse propagation as $\hat{\rm P}=\hat{\rm R}\hat{\rm S}\hat{\rm O}$, where $\hat{\rm S}$ is the operator for the SLM, $\hat{\rm O}$ represents other dominant operators %$\hat{\rm O}$ can be an individual or a combination of a few operators. 
and $\hat{\rm R}$ is a scaling operator acting to balance the net energy loss or gain from $\hat{\rm S} \hat{\rm O}$ \cite{wright2020mechanisms}. The final STML output $U(x, y, t)$ can be written as,

\begin{equation}
\begin{aligned}
U(x, y, t)=\lim _{k \rightarrow \infty}[\hat{\rm R} \hat{\rm S} \hat{\rm O}]^k U_0(x, y, t),
\label{eqn10}
\end{aligned}
\end{equation}
where $U_0(x, y, t)$ is the initial condition, and $k$ is the iterative number. When assuming the cavity is dominated by the SLM operator (omitting $\hat{\rm O}$), significant change in the output beam was simulated when varying $\delta\phi_r$ (Fig. \ref{fig3}d). When combining the SLM operator with either OS or SA, the simulated beam still changes when varying $\delta\phi_r$. Beam stability was simulated when including both OS and SA in $\hat{\rm O}$. Furthermore, the attractor dissection theory simulated beam in this case resembles the full model simulated beam resonably (see Fig. \ref{fig2}e and Fig. \ref{fig3}d). Therefore, it suggests that the SA and OS (contributing to xSF) underpin the observed spatiotemporal stabilizer. The presence of the SA operator in the spatiotemporal stabilization also distinguishes the STML state from the non-STML state.

\noindent{\bf Capability of the spatiotemporal stabilizer.} When $\delta\phi_r$ keeps increasing, the SLM operator can surpass the spatiotemporal stabilizer and cause the laser to lose STML. The maximum random phase $\delta\phi_r^{\rm max}$ that can be tolerated to retain the STML stability was measured to evaluate the capability of the spatiotemporal stabilizer.

For the 8 $\mu$m splicing offset, we observed $\delta\phi_r^{\rm max}$ increases with pulse energy (Fig. \ref{fig4}a). Each circled region in Fig. \ref{fig4}a represents an STML state attained by adjusting the waveplates and coupling from free space into MMFs, while dots in the circle were obtained by changing the pump power. Higher pulse energy means higher STML pulse peak power and lower SA loss (Supplementary Sec. 8). Thus, higher loss from SLM and a larger $\delta\phi_r$ can be tolerated. The increase of $\delta\phi_r^{\rm max}$ almost follows a linear relationship with the intracavity pulse energy. 

We then varied the splicing offset from 8 $\mu$m to 10 $\mu$m and 12 $\mu$m to further characterize this linear trend. The slope was observed to decrease with increasing splicing offset. Mode decomposition shows that the mode content condensed in the low-order mode (mode-1 to mode-5) decreases with increasing splicing offset (right panel Fig. \ref{fig4}a). More high-order modes were excited with a larger splicing offset and the measured beam profiles become more complex (see insets in Fig. \ref{fig4}a). These high-order modes are dissipated when coupling from the free space into the STIN gain fibre. Hence, the total cavity loss increases when more high-order modes are excited. Therefore, smaller splicing offset can result in a stronger spatiotemporal stabilization and a larger $\delta \phi_r^{\rm max}$ (larger loss from the SLM) can be tolerated. These measured trends on $\delta \phi_r^{\rm max}$ are consistent with numerical simulations using the full model, including the absolute slope value (Fig. \ref{fig4}b). The simulated beam profile also becomes more complex with a larger splicing offset. The dependence of $\delta \phi_r^{\rm max}$ on pulse energy and low-order mode proportion further confirm that OS and SA are responsible for resisting spatial phase perturbations.

\noindent{\bf Discussions.} We have shown that xSF and SA can act as a spatiotemporal stabilizer in STML lasers. By including spatial phase perturbations to the STML pulse beam using an SLM, it causes spatial-mode-dependent loss as well as walk-off between modal pulses. Despite significant changes in the STML pulse immediately after the SLM, the spatiotemporal stabilizer effectively corrects these perturbations to maintain consistent outputs across space, frequency, and time domains. %Our measurements and simulations constitute another building block to establish the fundamentals of STML. 
We also observed that higher low-order mode power condensation can enhance the resistance to spatial perturbations. However, more high-order mode excitation is favored to scale STML lasers towards higher pulse energy \cite{Wise_OL2022multimode,wright2017spatiotemporal,Bao_LSA2023spatiotemporal}. Fortunately, higher pulse energy itself enhances the capability to resist spatial perturbations. Hence, simultaneous design considerations for high-order-mode excitation and increased pulse energy are essential to achieve greater spatiotemporal stability in high energy STML lasers. This requirement is rooted in the combined contributions from SA and xSF to the spatiotemporal stabilizer. %\textcolor{red}{we often want to control physical systems and change their properties, but often these systems are delicate. Therefore we need to be intelligent with how, when, and how fast we perturb these systems. STML laser, being so high-dimensional, is a nice little microcosm of that.} 
In parallel with pulse energy scaling, beam profile engineering is also another important goal for STML lasers \cite{fu2018several,Bao_LSA2023spatiotemporal}. Including an SLM in the cavity has been regarded as a potential solution to this task. Our work reminds us that the dominant operators should include the SLM operator to efficiently engineer the output beams. Another important consideration is to find the optimal positions to place the SLM and derive the output. For example, pulses right after the SLM do change, although STML pulses after propagating through the cavity have a relatively steady output. Therefore, by designing laser configuration appropriately, intracavity mode engineering using an SLM is still possible, despite the existence of the spatiotemporal stablizer. Furthermore, the SLM output was coupled into the gain STIN MMF supporting a small mode number in the current laser. By adding a GRIN MMF before the gain MMF, the SLM output can be coupled into an MMF supporting much more spatial modes. It can be interesting to investigate the spatiotemporal stabilization dynamics in such a laser configuration.

\vspace{3 mm}
\noindent{\bf Methods}

{\small
\noindent \textbf{CNN-SPGD-based Mode decomposition.} The convolution neural network (CNN) approach has been widely used for mode decomposition for MMF outputs \cite{an2020deep,bruning2013comparative,an2019learning,an2019numerical}. Typically, MMFs comprising less than 10 modes are used for mode decomposition \cite{an2019numerical}. For a larger mode number, it has been proposed that CNN-based mode decomposition can be used together with SPGD algorithm to improve the accuracy \cite{bruning2013comparative,an2019numerical}. Following this proposed scheme, we used the CNN-SPGD approach to decomposite the STML beam. Specifically, we used the VGG-16 CNN model \cite{simonyan2014very} for initial mode decomposition and used the CNN output as the initial condition for the SPGD optimization \cite{an2019numerical} to yield the final output.  

We used the lowest 20 modes of the passive MMF to train the CNN. Although the passive GRIN MMF can support about 100 modes, this simplification is still reasonable, as the splicing offset between the gain MMF and passive MMF was not very large and very high-order modes were not excited experimentally. We did not used the full image, but selected a region of the measured beam for decomposition. The used beam region is no larger than the theoretically calculated superposition of the wanted 20 modes.

The flowchart of our mode decomposition method is illustrated in Supplementary Fig. S4. A plethora of dataset of beam profiles and their corresponding mode content is generated by superimposing the considered spatial modes to train the CNN. The input beam profile undergo a series of transformations, including convolution and pooling, as they traverse the various layers of the CNN to attain an output column vector. The loss function of the CNN is defined as the mean squared error (MSE) between the output vector and the given mode content vector. The training process involves the iterative use of the back propagation algorithm to refine the performance of the CNN by reducing the loss function progressively. When the MSE reaches a given threshold, we deemed the training has converged. Subsequently, the trained CNN was used to decomposite various numerically generated beam profiles to assess whether a sufficient accuracy has been attained. After that, the trained CNN was used for mode decomposition of the measured beams. Then, the retrieved mode content from the CNN serves as the initial condition for the SPGD algorithm to further minimize the MSE to yield the final output.

%\vspace{1mm}
%\noindent \textbf{Simulation of spatiotemporal stability in STML.} 

}

\vspace{3 mm}
\noindent \textbf{Data Availability.}
The data that supports the plots within this paper and other findings is available.

\vspace{1 mm}

\noindent \textbf{Code Availability}
The code that supports findings of this study are available from the corresponding author upon reasonable request.

%\vspace{1 mm}
\noindent \textbf{Acknowledgments.}
This work is supported by the National Natural Science Foundation of China (62375150, 62175127, 62250071, 62375024), the National Key R\&D Program of China (2021YFB2801200), the Tsinghua-Toyota Joint Research Fund, and the Tsinghua University Initiative Scientific Research Program (20221080069).

\vspace{1 mm}
\noindent\textbf{Author Contributions.} G.C., C.Y. and C.B. conceived the project. G.C. led the experiment with assistance from B.C.; G.C. led the simulations and mode decomposition with assistance from C.W., Z.J. and X.S. The project was supervised by C.B.  
\vspace{1 mm}

\noindent \textbf{Competing Interests.} The authors declare no competing interests.

%\vspace{1 mm}

%\noindent \textbf{Author Information} Correspondence and requests for materials should be addressed to C.Z. (chijie@tsinghua.edu.cn), C.B. (cbao@tsinghua.edu.cn) and R.Z. (zengrong@tsinghua.edu.cn).

% \bibliography{main.bib}
\bibliography{ref}

% \printbibliography

% % remove header but keeping the page num
% \thispagestyle{fancy}
% \fancyhead{}    % remove header
% \fancyfoot{}    % remove footer
% \renewcommand{\headrulewidth}{0pt}  % remove the line
% \fancyhead[R]{\thepage}

\end{document}